\documentclass[twocolumn]{aastex631}

\def\h2o  {H$_2$O}

\begin{document}

\title{The Detection of H$_{2}$O Maser Emission from mid-IR Red Galaxies  }

\author{C. Y. Kuo}
\affiliation{Physics Department, National Sun Yat-Sen University, No. 70, Lien-Hai Rd, Kaosiung City 80424, Taiwan, R.O.C  } 
\affiliation{Academia Sinica Institute of Astronomy and Astrophysics, P.O. Box 23-141, Taipei 10617, Taiwan, R.O.C.  }

\author{C. Y. Tai}
\affiliation{  Graduate Institute of Astronomy, National Central University, \\
No. 300, Zhongda Rd., Zhongli Dist., Taoyuan City, 320317, Taiwan (R.O.C.)}
\affiliation{Department of Applied Physics, Kaosiung University, No. 700, Gaoxiongdaxue Rd., Nanzi Dist., Kaohsiung City 811726, Taiwan, R.O.C. }

\author{A. Constantin}
\affiliation{ Department of Physics and Astronomy, James Madison University, Harrisonburg, VA 22807, USA }

\author{J. A. Braatz}
\affiliation{National Radio Astronomy Observatory, 520 Edgemont Road, Charlottesville, VA 22903, USA   }

\author{H. H. Chung}
\affiliation{Physics Department, National Sun Yat-Sen University, No. 70, Lien-Hai Rd, Kaosiung City 80424, Taiwan, R.O.C }

\author{B. Y. Chen}
\affiliation{Physics Department, National Sun Yat-Sen University, No. 70, Lien-Hai Rd, Kaosiung City 80424, Taiwan, R.O.C }

\author{ D. W. Pesce}
\affiliation{ Black Hole Initiative at Harvard University, 20 Garden Street, Cambridge, MA 02138, USA   }
\affiliation{ Harvard-Smithsonian Center for Astrophysics, 60 Garden Street, Cambridge, MA 02138, USA  }

\author{C. M. V. Impellizzeri}
\affiliation{Leiden Observatory, Leiden University, PO Box 9513, 2300 RA Leiden, The Netherlands }

\author{F. Gao}
\affiliation{ Max Planck Institute for Radio Astronomy, auf dem H$\ddot{u}$gel 69, 53121 Bonn, Germany}

\author{ Y.-Y. Chang}
\affiliation{Department of Physics, National Chung Hsing University, No. 145, Xingda Rd., South Dist., Taichung City 402202, , Taiwan (R.O.C.)}



\begin{abstract}


We report the detection of H$_{2}$O maser emission in 4 out of 77 (5.2\%) mid-IR red galaxies that meet the color criteria of $W1-W2 > 0.5$ and $W1-W4 > 7$ and are classified as Type-2 AGNs based on optical, near-IR, and mid-IR spectral energy distribution (SED) fitting. Here, $W1$, $W2$, and $W4$ represent the IR magnitudes at 3.4, 4.6, and 22 $\mu$m, respectively, as measured by the Wide-field Infrared Survey Explorer.  Three of the four newly identified maser galaxies are classified as either Seyfert 2 or LINER systems, but none are disk maser systems. Our analysis indicates that AGN identifications based solely on SED fitting are unreliable, resulting in an unexpectedly low detection rate. By restricting our sample to optically classified Type 2 AGNs that satisfy the mid-IR color criteria, we achieve a maser detection rate of $\sim$13$-$18\%, aligning with previous predictions for mid-IR red sources. These selection criteria are the most effective to date for facilitating new maser detections, particularly in light of the recent identification of additional Type 2 AGNs identified from ongoing galaxy and AGN surveys.

\end{abstract}

\keywords{masers --- galaxies: Seyfert --- infrared: galaxies --- galaxies: active --- (galaxies:) quasars: supermassive black holes }


\section{Introduction}

Since the discovery of the subparsec Keplerian H$_{2}$O maser disk at the center of the archetypal maser galaxy NGC 4258 \citep[e.g.][]{miyoshi95, hmg99, argon07}, there has been a significant interest in searching for similar maser systems over the past three decades, because 
they offer valuable insights for various astrophysical applications. For instance, the Megamaser Cosmology Project (MCP; \citealt{reid09, bra10}) has demonstrated that 22 GHz H$_{2}$O disk megamasers enable a direct determination of the Hubble Constant ($H_{0}$) independent of standard candles and facilitate precise measurements of supermassive black hole (BH) masses ($M_{\rm BH}$) by modeling gas distribution and kinematics traced by H$_{2}$O maser emission from Keplerian disk maser systems, such as NGC 4258. The recent $H_0$ measurement (i.e. $H_0 = 73.9 \pm 3.0$ km s$^{-1}$ Mpc$^{-1}$; \citealt{dom20}) obtained using this ``megamaser technique" offers crucial evidence in support of the apparent tension between early- and late-universe measurements of $H_0$ \citep[see, e.g.][]{abdalla22, verde19}.

For the purpose of identifying H$_{2}$O maser disks for $H_0$ and BH mass measurements, the MCP initiated a multi-year campaign two decades ago. This effort involved surveying for 22 GHz water megamaser emission in $\gtrsim$4800 galaxies with active galactic nuclei, monitoring the spectral evolution of the detected megamasers, and conducting sub-milliarcsecond imaging of those systems with Very Long Baseline Interferometry (VLBI). While the MCP maser survey has primarily focused on searching for maser disks in galaxies with narrow emission-line activity in their centers (i.e. type 2 AGNs, including both Seyfert 2 galaxies and some Low Ionization Nuclear Emission Regions, or LINERs), which are expected to host a significantly higher fraction of H$_{2}$O megamasers compared with type 1 AGNs \citep[see details in][]{kuo18}, the chances of finding these megamasers remained relatively low. Overall, the detection rate ($R_{\rm maser}$) of any water maser emission at 22 GHz is $R_{\rm maser}\sim$3\%, and the likelihood of finding those exhibiting a disk-like configuration is $R_{\rm disk}\sim$1\%. Such low detection rates significantly limit the applicability of H$_{2}$O megamaser activity in astronomy, particularly in obtaining a large number of accurate black hole masses.

To improve the detection rate of masers, several efforts have been made to identify observational signatures that would indicate a high likelihood of detectable masers in various types of galaxies \citep[e.g.][]{zhu11,constantin12, zhang12, liu17, kuo20}. For example, in \citet{kuo18}, we analyzed the $\sim 4800$ galaxies in the MCP maser survey and found that H$_{2}$O megamaser emission tends to be associated with strong emission in all {\it Wide-field Infrared Survey Explorer (WISE)} mid-IR wavelengths, with the strongest enhancement in the $W4$ band, at 22 $\micron$; the maser detection rates can be significantly enhanced by surveying type 2 AGNs with redder mid-IR colors, with the highest detection rate ($R_{\rm maser} = 18.2\pm 2.5$\%) being obtained by selecting galaxies with $WISE$ colors satisfying $W1-W2 > 0.5$ \& $W1-W4 > 7$, where $W1$ and $W2$ are the WISE magnitudes at 3.4 $\micron$ and 4.6 $\micron$, respectively. The association of the high $R_{\rm maser}$ with this particular set of mid-IR color selections is consistent with the finding that H$_{2}$O megamasers are detected preferentially in heavily obscured AGNs \citep{kuo20}, where dust heating leads to enhanced mid-IR emission.  

In this paper, we report four new detections of H$_{2}$O megamasers from a 22 GHz survey with the Green Bank Telescope (GBT).  Our aim was to investigate whether surveying galaxies with the specific mid-IR properties predicted by \citet{kuo18} can indeed lead to a significant enhancement in the megamaser detection rate. In Section \ref{sec:sample}, we present our sample of galaxies and describe the GBT observations as well as the data reduction process. Section \ref{sec:results} presents the new detections and explores the properties of these megamaser systems. We discuss in Section \ref{sec:discussion} the significance of these detections, along with the impact of the uncertainty in the AGN classification on the maser detection rate, via a comparison with the expected rates from the MCP survey. We present the conclusions of this survey and study in Section \ref{sec:conclusion}.

\section{The Sample, Observations, and Data Reduction} \label{sec:sample}

\subsection{The Sample of Red mid-IR Galaxies} \label{sec:sample_selection}

We selected candidate galaxies for our survey from five well-known catalogs: 2df \citep{colless2001}, 6df \citep{jones2009}, 2MRS \citep{huchra2012}, RC3 \citep{RC3_1991}, and Galaxy Zoo \citep{galaxyzoo13}. We required the galaxies to have declinations greater than $-$30$^{\circ}$ and mid-IR photometry at all four WISE bands, 3.4, 4.6, 12, and 22 $\micron$ ($W1, W2, W3, W4$). Additionally, we required the galaxies to have spectroscopic redshifts $z\lesssim 0.05$, the same redshift limit of the MCP sample from which we derived the mid-IR selection criteria in \citet{kuo18}. This selection resulted in a total of 47,133 galaxies.

We cross-matched our collected sample with the MCP catalog of galaxies surveyed with the GBT, which contains 4860 sources (\url{https://www.gb.nrao.edu/~jbraatz/H2O/sum_dir_sort.txt}) and removed matches. We then collected the profile-fit magnitudes $W1, W2, W3, W4$ for our sample that would capture all of the AGN flux while excluding mid-IR emission from starforming activity outside of the point-spread function. We applied the criteria $W1-W2 > 0.5$ and $W1-W4 > 7$, which are anticipated to boost the maser detection rate significantly \citep{kuo18}. Galaxies with such red mid-IR colors are rare in the local Universe ($\sim$6\% of the MCP sample), so imposing the mid-IR color cuts significantly reduced the sample to 171 galaxies. 
Given their red colors in the mid-IR, the majority of these galaxies are expected to contain AGN components \citep[e.g.][]{satyapal2014}.

\begin{figure}[ht]
\begin{center} 
\vspace*{0.5 cm} 
\hspace*{-0.3 cm} 
\includegraphics[trim = 30 10 0 0, clip, scale=0.21]{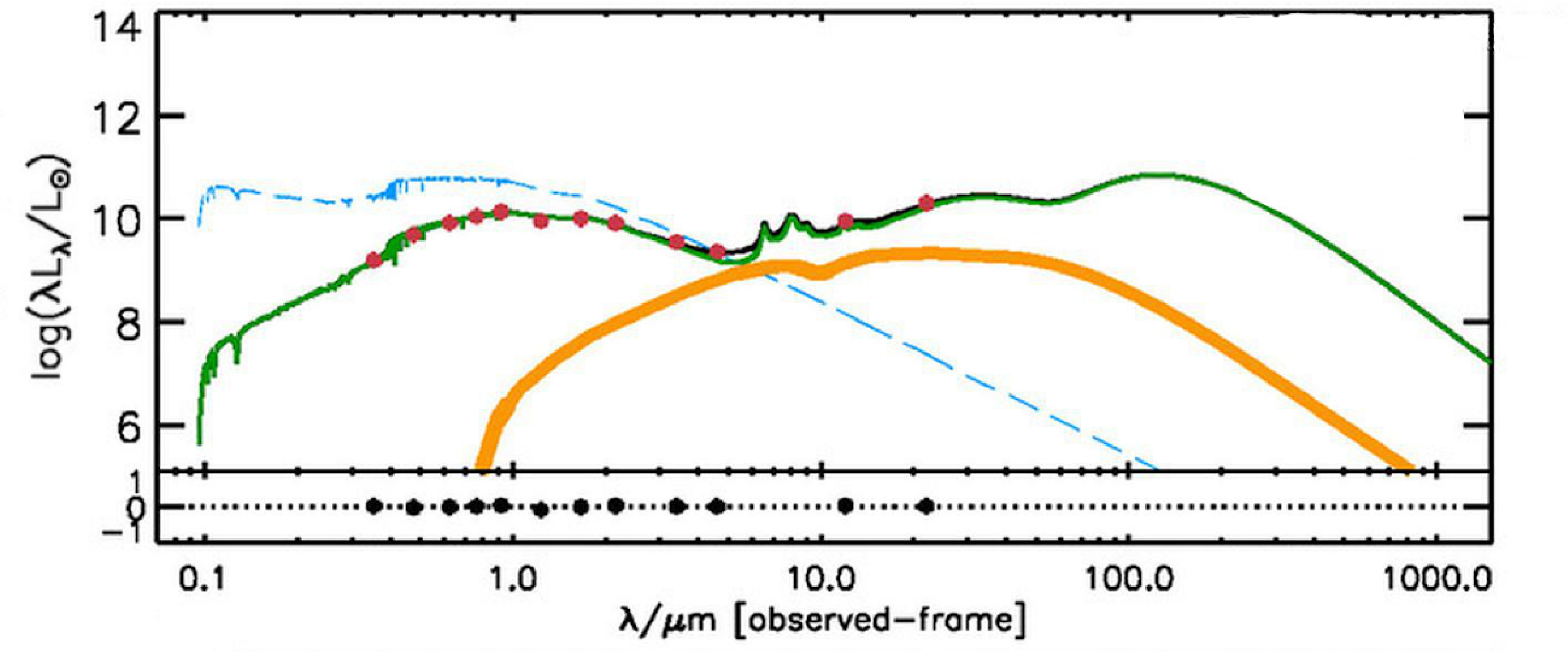}
\vspace*{0 cm} 
\caption{An example of the SED fitting for our sample galaxies. The red dots are photometric data at the optical, near-IR, and mid-IR wavelengths \citep[see details in][]{kuo20} for the source TGN278Z058. The green lines represent emission from the host galaxy and the blue lines show the SED of the {\it unattenuated} stellar continuum. The contribution from the Seyfert 2 nucleus in the host galaxy is shown by the orange line and the black line indicates the best-fit SED model. }
\label{fig:SED_fitting}
\end{center} 
\end{figure} 

Since the imposed mid-IR criteria were derived from the catalog of the MCP maser survey, which primarily targeted type 2 AGNs, these criteria are most suitable for similar sources. However, we note that the majority of the 171 sources satisfying the criteria do not have the available spectroscopic data required to identify narrow emission line AGNs based on the BPT line diagnostic diagram \citep[e.g.][]{baldwin81}. Thus, to identify type 2 AGNs for our maser survey, we fit the spectral energy distributions (SEDs) to our galaxies using their optical, near-IR, and mid-IR photometry \citep[see details in][]{kuo20} by adopting the widely used SED fitting code MAGPHYS \citep[e.g.][]{chang17}. 
Our fitting results (see an example in Figure \ref{fig:SED_fitting}) indicate type 1 AGN for 79 of the galaxies, while 78 galaxies show SEDs consistent with Seyfert 2s.  The remaining 14 galaxies in our sample could not be fit adequately with any of the available models, and we thus excluded them from the survey since we can't assign any clear AGN classification to these sources. To optimize the detection rate, our survey mainly focused on the galaxies classified as Seyfert 2s by MAGPHYS.  We discuss the impact of the uncertainty in the SED-based AGN classification on the expected maser detection rate in Section \ref{sec:discussion}. The basic properties of the 78 SED-based Seyfert 2s are shown in Table \ref{tab:1}.

\subsection{GBT Observations and Data Reduction} \label{gbtdata}

 We used the Robert C. Byrd Green Bank Telescope (GBT) to search for 22 GHz H$_{2}$O maser emission in our mid-IR red galaxy sample during February and March 2019 (Project ID: GBT19A-340). We successfully observed 77 of the 78 galaxies in our sample. We observed in dual polarization using the KFPA receiver (18 -- 26.5 GHz) and the VEGAS backend (mode 4\footnote{ \url{https://www.gb.nrao.edu/vegas/}}). This setup provides a bandwidth of 187.5 MHz and a spectral resolution of 5.7 kHz, enabling a total spectral coverage of approximately 2600 km~s$^{-1}$ and a velocity resolution $\Delta v$ of about 0.08 km~s$^{-1}$. Here, we select a velocity resolution that is substantially smaller than the typical maser linewidth (i.e. $\sim$2 km~s$^{-1}$) in order to identify channels affected by Radio Frequency Interference (RFI), which is often narrow in frequency (i.e. velocity width $\sim$ 0.1$-$0.2 km~s$^{-1}$), and minimize its impact on the maser spectrum.
 
 We determined the antenna temperature $T_{\rm a}$ of the target sources through dual-beam position-switched observations using the standard {\it Nod} procedure, which involves taking on-source "signal" and off-source "reference" spectra simultaneously with two beams. For most of the sample, each target was observed with two pairs of 2.5-minute scans, where beam 1 and beam 2 interchangeably collected signal and reference data. For four galaxies (TGN215Z110, CGCG481$-$011, IRAS02438$+$2122, CGCG 147$-$020), we observed for additional time (8--14 scans) to increase sensitivity.

Data reduction was performed using the {\it GBTIDL} analysis tool \footnote{\url{https://www.gb.nrao.edu/GBT/DA/gbtidl/users_guide/}}. The {\it getnod} procedure was employed to extract scans from the raw data file and to apply basic calibration, given by $T_{\rm a} = T^{\rm ref}_{\rm sys}\times[sig(\nu)-ref(\nu)]/ref(\nu)$, where $T^{\rm ref}_{\rm sys}$ is the system temperature at the reference position, and $sig(\nu)$ and $ref(\nu)$ represent the signal and reference spectra, respectively. To enhance the signal-to-noise ratio (SNR), we set the {\it getnod} parameter {\it smthoff} to 16, applying boxcar smoothing with a convolution size of 16 channels to the reference spectrum, which improved sensitivity by approximately 30\%. After removing channels affected by RFI, we averaged the spectra and subtracted the spectral baseline to obtain the fully calibrated spectra. The 1-$\sigma$ rms noise of the unsmoothed spectrum (i.e. $\Delta v \sim$ 0.08 km~s$^{-1}$) for each source is listed in Column (8) of Table \ref{tab:1}.

\startlongtable
\begin{deluxetable*}{llrrrrccc} 
\tablenum{1}
\tablewidth{0 pt} 
\tablecaption{Basic Properties of the 78 SED-based Seyfert 2 red galaxies in our Sample}
\tablehead{ 
\colhead{Galaxy} & \colhead{Catalog} & \colhead{RA} & \colhead{DEC} & \colhead{Redshift} & \colhead{Velocity} & \colhead{Distance} & \colhead{1-$\sigma$ rms} & \colhead{Activity}  \\
\colhead{Name}    &  \colhead{}  &  \colhead{} & \colhead{} & \colhead{ } & \colhead{(km~s$^{-1}$)} & \colhead{(Mpc)} & \colhead{ (mJy) } & \colhead{Type } 
}     
\startdata 
 TGS 807Z359           &   2df    & 00:05:23.35 & $-$08:10:12.10 & 0.02993  & 8972     & 131      & 15.7          & ---     \\   
  UGC 265              &   2MRS   & 00:27:14.27 & $+$20:04:19.99 & 0.01866  & 5593     & 81       & 8.9           & AGN? \\ 
 2MASX J00544183-1714416  &  6df  & 00:54:41.82 & $-$17:14:41.60 & 0.04824  & 14462    & 214      & 9.8           &  EmG \\
  2MASX J01112119+2755430 &  2MRS & 01:11:21.21 & $+$27:55:42.96 & 0.03334  & 995      & 146      & 9.6           & ---     \\        
  CGCG 481-011         &  2MRS   & 01:35:10.58 & $+$21:54:04.54 & 0.04732  & 14186    & 210      & 6.2           & ---  \\      
  NGC 0814             &  2MRS   & 02:10:37.63 & $-$15:46:24.92 & 0.00539  & 1616     & 23       & 22.0          & AGN?\\ 
  IRAS F02393$-$1752   &  6df    & 02:41:40.90 & $-$17:39:28.90 & 0.03438  & 10306    & 151      & 14.8         & Sy2   \\        
  NGC 1064             & 6df    & 02:42:23.55 & $-$09:21:44.10 & 0.03145  & 9428     & 138      & 10.8          & Sy1   \\        
  IRAS 02438+2122      &  2MRS  & 02:46:39.19 & $+$21:35:09.38 & 0.02331  & 6987     & 102      & 5.7           & --- \\       
  NGC 1377             &  2MRS   & 03:36:39.08 & $-$20:54:08.14 & 0.0059   & 1768     & 25       & 11.1          & ---  \\   
 2MASXJ04023209+6020377  & 2MRS   &  04:02:32.09 &  +60:20:38.08 &  0.0300  &  8997   &  132    & ---             & --- \\
  2MASX J04250311-2521201  &  6df  & 04:25:03.12 & $-$25:21:20.20 & 0.04183  & 12542    & 185      & 12.3          & Sy2  \\         
  UGC 03094           &  2MRS   & 04:35:33.83 & $+$19:10:18.16 & 0.02471  & 7408     & 108      & 10.6          & --- \\       
  LEDA 16061          & 6df   & 04:48:35.27 & $-$04:49:10.50 & 0.0159   & 4768     & 69       & 18.0          & EmG  \\         
  2MASX J05235546+4842509 &  2MRS & 05:23:55.48 & $+$48:42:51.23 & 0.03359  & 10070    & 148      & 8.7           & ---  \\      
  LEDA 966552          &  2MRS   & 05:52:33.21 & $-$11:22:29.17 & 0.02186  & 6553     & 95       & 12.8          & Sy2 \\          
  ESO 555-G022       &   2MRS   & 06:01:07.97 & $-$21:44:19.00 & 0.00583  & 1748     & 25       & 16.0          & --- \\          
  ESO 557-2          &   2MRS  & 06:31:47.22 & $-$17:37:17.33 & 0.02118  & 6351     & 92       & 13.2          & EmG  \\
  CGCG 147-020       &   2MRS  & 07:25:37.25 & $+$29:57:14.76 & 0.01885  & 5650     & 82       & 4.4           & Sy2  \\         
  CGCG 058-009       &   2MRS  & 07:35:43.43 & $+$11:42:35.24 & 0.01625  & 4873     & 70       & 14.3          & ---  \\           
  2MASX J08353838-0114072 &  6df  & 08:35:38.40 & $-$01:14:07.20 & 0.04378  & 13123    & 194      & 12.3          & --- \\       
  2MASS J09082215+3206467  & GZ & 09:08:22.15 & $+$32:06:46.65 & 0.04961  & 14873    & 221      & 13.9          & Sy2 \\          
  UGC 04881NED01        &  2MRS   & 09:15:55.50 & $+$44:19:57.76 & 0.03978  & 11927    & 176      & 11.8          & --- \\  
  2MASX J09491796+1750444 & GZ & 09:49:17.94 & $+$17:50:45.03 & 0.0498   & 14930    & 221      & 11.6          & AGN \\   
  TGN 215Z110             & 2df  & 09:54:27.04 & $-$03:12:58.70 & 0.04855  & 14556    & 216      & 9.4           & ---  \\      
  TGN 353Z089          &  2df    & 10:03:01.37 & $-$00:09:55.60 & 0.04471  & 13403    & 198      & 9.2           & ---  \\      
  TGN 289Z049          &   2df    & 10:17:09.12 & $-$01:47:38.30 & 0.04552  & 13645    & 202      & 9.4           & rG  \\      
  LEDA 860076          &  6df   & 10:26:50.04 & $-$19:04:32.00 & 0.03027  & 9074     & 133      & 11.0          & Sy2     \\      
  SDSS J103023.34+282059.0 & GZ & 10:30:23.34 & $+$28:20:59.04 & 0.03732  & 11188    & 164      & 10.8          & ---  \\      
  NVSS J103317+162425   & GZ   & 10:33:17.11 & $+$16:24:25.45 & 0.04505  & 13504    & 200      & 10.8          & rG  \\      
  TGN 229Z166          &  2df   & 10:58:35.36 & $-$02:35:53.10 & 0.03623  & 10862    & 160      & 12.7          & rG  \\      
  LEDA 33083         &  2MRS     & 10:59:18.14 & $+$24:32:34.44 & 0.0431   & 12921    & 191      & 9.0           & Liner   \\      
  FIRST J111106.5+060102 & GZ  & 11:11:06.60 & $+$06:01:01.95 & 0.04376  & 13119    & 194      & 13.0          & rG \\  
  LEDA 822452         &  6df     & 11:20:06.98 & $-$21:52:47.00 & 0.02734  & 8195     & 120      & 11.4          & --- \\       
  TGN 234Z011         &  2MRS    & 11:21:12.23 & $-$02:59:02.60 & 0.02482  & 7441     & 108      & 14.0          & Sy2  \\         
  IRAS 11215-2806     &  2MRS     & 11:24:02.73 & $-$28:23:15.50 & 0.01376  & 4125     & 60       & 13.9          & Sy2  \\         
  NVSS J113639+173836  &  2MRS    & 11:36:39.94 & $+$17:38:36.49 & 0.02726  & 8171     & 119      & 14.3          & rG \\  
  TGN377Z135        &   2df      & 11:45:40.24 & $-$00:34:15.40 & 0.02805  & 8410     & 123      & 14.3          & ---   \\     
  IC 0737          &  2MRS    & 11:48:27.53 & $+$12:43:38.32 & 0.01371  & 4109     & 59       & 13.6          & ---   \\     
  TGN 242Z154      &  2df     & 11:51:33.06 & $-$02:22:23.10 & 0.00357  & 1071     & 15       & 12.5          & ---   \\     
  TGN 311Z206      &  2df     & 11:52:47.51 & $-$00:40:08.50 & 0.00455  & 1365     & 20       & 11.6          & Sy1      \\     
  A1203+31B      &   RC3   & 12:05:45.50 & $+$31:03:32.04 & 0.02297  & 6886     & 100      & 11.0          & rG   \\     
  SDSS J120628.49+633747.2  &  GZ & 12:06:28.49 & $+$63:37:47.28 & 0.03955  & 11858    & 175      & 16.8          & ---   \\     
  TGN121Z264       &   2MRS    & 12:14:51.26 & $-$03:29:22.80 & 0.03377  & 10122    & 148      & 9.0           & Sy1      \\     
  UGC 7531         &  RC3    & 12:26:13.39 & $-$01:18:18.00 & 0.006    & 1800     & 26       & 8.7           & ---        \\     
  SDSS J125224.34+090934.6  &  GZ & 12:52:24.35 & $+$09:09:34.63 & 0.02594  & 7776     & 113      & 8.3           & ---   \\     
  TGN 130Z145        &  2df    & 12:56:15.84 & $-$03:27:22.60 & 0.04498  & 13484    & 199      & 8.6           & ---   \\     
  NVSS J130150+042001  &  2MRS   & 13:01:50.28 & $+$04:20:00.49 & 0.03736  & 11200    & 165      & 8.6           & rG \\  
  2MASX J13042219+3615428  & GZ  & 13:04:22.19 & $+$36:15:43.15 & 0.04426  & 13269    & 196      & 8.0           & Sy2    \\       
  LEDA 1021744        &  6df     & 13:11:16.67 & $-$07:16:19.50 & 0.02262  & 6782     & 99       & 9.3           & EmG    \\
  TGN 135Z219         &  GZ     & 13:14:27.01 & $-$03:36:37.00 & 0.04261  & 12774    & 188      & 8.5           & Sy1    \\       
  SDSS J131654.33+465756.6 & GZ & 13:16:54.33 & $+$46:57:56.69 & 0.04955  & 14854    & 220      & 7.7           & --- \\       
  IC 0883         &   2MRS   & 13:20:35.35 & $+$34:08:21.91 & 0.02306  & 6912     & 101      & 8.3           & SBG \\   
  A1326+44        &   RC3    & 13:28:44.40 & $+$43:55:51.96 & 0.02824  & 8467     & 124      & 8.6           & ---     \\        
  ESO 509-IG066NED01  &  2MRS  & 13:34:39.64 & $-$23:26:47.54 & 0.03435  & 10299    & 151      & 8.8           & Sy2   \\        
  IC 0910          &  2MRS      & 13:41:07.84 & $+$23:16:55.09 & 0.02713  & 8133     & 119      & 10.5          & Liner \\        
  TGN 338Z025      &   2df    & 13:51:15.79 & $+$00:23:27.90 & 0.02991  & 8966     & 131      & 8.7           & Liner \\        
  TGN 406Z112      &   2df   & 13:58:17.39 & $+$01:08:46.20 & 0.03474  & 10415    & 153      & 8.5           & ---\\        
  IRAS 13559-1553  &   2MRS  & 13:58:40.56 & $-$16:08:25.80 & 0.03628  & 10876    & 160      & 9.9           & Sy2  \\         
  TGN 206Z075    &  2df    & 14:14:38.30 & $-$02:59:45.90 & 0.04635  & 13896    & 206      & 9.3           & --- \\       
  NGC 5610       &  2MRS    & 14:24:22.94 & $+$24:36:51.26 & 0.01689  & 5063     & 73       & 9.6           & Sy2  \\         
  TGN 278Z058     &  2df    & 14:29:48.72 & $-$01:10:08.00 & 0.03016  & 9041     & 132      & 8.7           & --- \\        
  2MASS J14343852+2304432  &  GZ & 14:34:38.51 & $+$23:04:43.24 & 0.03737  & 11204    & 165      & 8.0           & Sy2  \\         
  SDSS J144535.33+500540.0 &  GZ & 14:45:35.33 & $+$50:05:40.04 & 0.03076  & 9222     & 135      & 9.6           & --- \\       
  2MASX J15055739+0520157  &  GZ & 15:05:57.40 & $+$05:20:15.67 & 0.03717  & 11144    & 164      & 9.8           & --- \\       
  2MASX J15062421+3225507  &  GZ  & 15:06:24.24 & $+$32:25:51.03 & 0.04339  & 13007    & 192      & 8.1           & Liner \\        
  2MASX J15281225+3719168  &  GZ  & 15:28:12.23 & $+$37:19:16.47 & 0.03253  & 9753     & 143      & 8.1           & Sy2   \\        
  NVSS J152814+025237      &  GZ  & 15:28:14.69 & $+$02:52:36.46 & 0.0387   & 11601    & 171      & 9.9           & rG \\ 
  2MASX J15471040+2706272  &  GZ  & 15:47:10.44 & $+$27:06:27.74 & 0.03277  & 9824     & 144      & 7.8           & ---    \\    
  SDSS J160127.94+090654.6 &  GZ   & 16:01:27.95 & $+$09:06:54.60 & 0.03427  & 10274    & 151      & 8.5           & ---    \\    
  SDSS J160524.19+480554.5 &  GZ & 16:05:24.20 & $+$48:05:54.52 & 0.04421  & 13254    & 196      & 7.7           & ---    \\    
  SDSS J161243.21+124505.6 &  GZ & 16:12:43.21 & $+$12:45:05.60 & 0.03473  & 10412    & 153      & 8.5           & ---   \\     
  MCG +04-39-009     &  GZ     & 16:30:07.64 & $+$21:50:47.82 & 0.03662  & 10979    & 161      & 7.8           & Sy2      \\     
  LEDA 90310      &   2MRS   & 18:52:22.44 & $-$29:36:20.56 & 0.0424   & 12711    & 188      & 15.3          & Sy1      \\     
  LEDA 855228     &   2MRS   & 20:50:03.77 & $-$19:26:20.51 & 0.02714  & 8135     & 119      & 23.3          & EmG      \\     
  ESO 530-G025    &  2MRS       & 21:14:09.48 & $-$26:03:37.94 & 0.02686  & 8051     & 118      & 16.9          & ---   \\     
  Z379-24       &  2MRS     & 23:03:41.23 & $+$01:02:36.67 & 0.04205  & 12605    & 186      & 9.1           & ---      \\     
  Z381-51       &   2MRS     & 23:48:41.72 & $+$02:14:23.10 & 0.03067  & 9194     & 134      & 12.9          & AGN      \\ 
\enddata
\tablecomments{Column (1): Galaxy name; Column (2): The catalog from which the galaxy was selected; Columns (3) \& (4): Right Ascension (RA) and Declination (DEC) of the galaxy, respectively; Column (5): Redshift of the source; Column (6): Heliocentric recession velocity of the galaxy evaluated using the “optical” velocity convention; Column (7): Luminosity distance of the source, calculated based on the standard $\Lambda$CDM cosmology, with $\Omega_{\rm matter} = 0.3089$, $\Omega_{\rm vacuum} = 0.6911$, and $H_{0} = 70$ km~s$^{-1}$ Mpc$^{-1}$; Column (8): 1-$\sigma$ RMS noise in units of mJy for the unsmoothed spectrum obtained from the data reduction process, which has a spectral resolution of 0.08 km~s$^{-1}$; Column (9): Activity type of the galaxy as listed in SIMBAD. Here, AGN? indicates an AGN candidate, EmG refers to an emission-line galaxy dominated by star formation, Sy1 denotes a Seyfert 1 galaxy, Sy2 denotes a Seyfert 2 galaxy, SBG indicates a starburst galaxy, rG stands for a radio galaxy, Liner refers to an AGN hosting a low-ionization nuclear emission-line region, and AGN refers to an active galactic nucleus without a precisely known subtype.}
\label{tab:1}
\end{deluxetable*}

\section{Results} \label{sec:results}

\begin{figure*}[ht]
\begin{center} 
\vspace*{0 cm} 
\hspace*{0 cm} 
\includegraphics[trim = 30 320 30 0, clip,angle=0, scale=0.6]{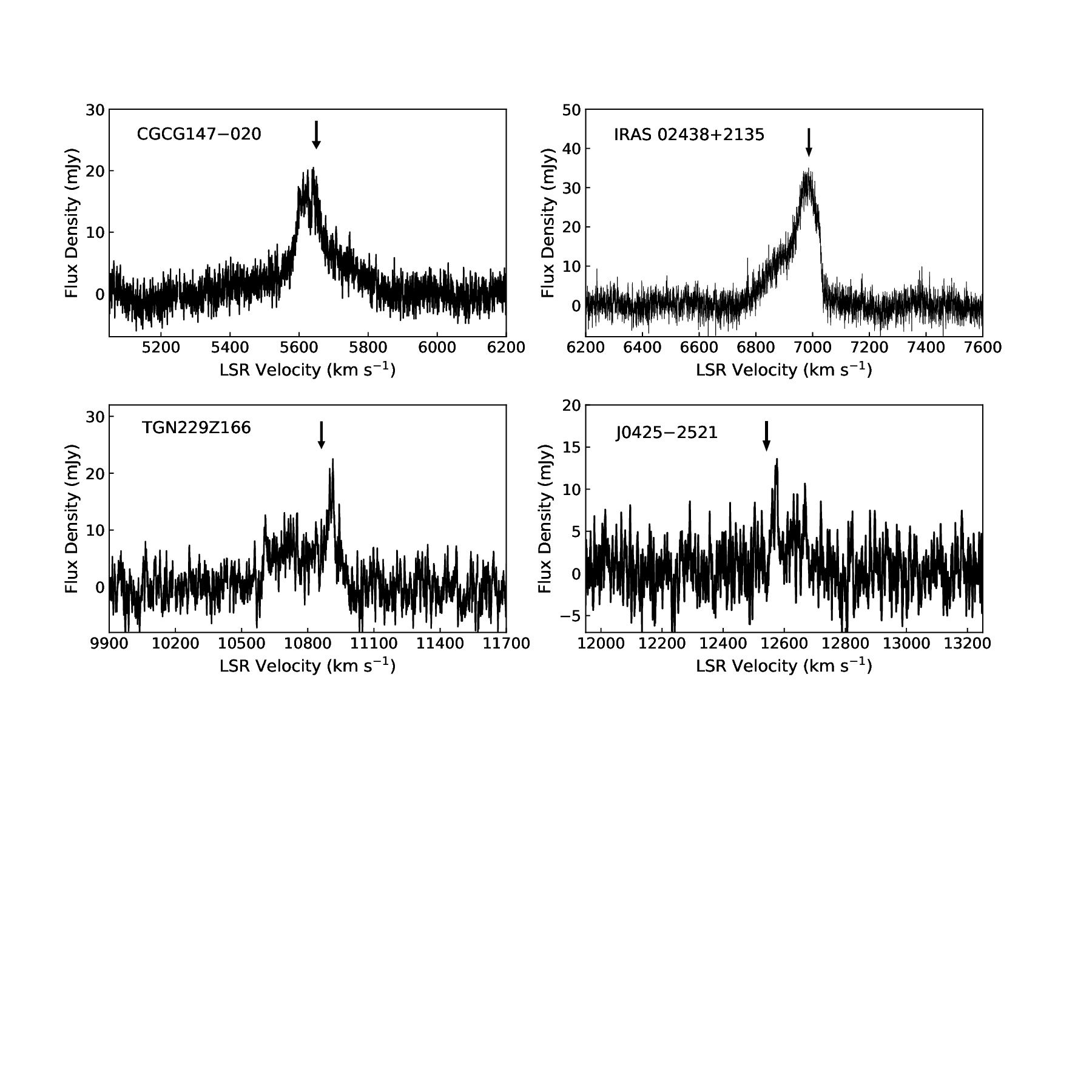}
\vspace*{0.0 cm} 
\caption{Spectra of the four new megamaser detections. The x-axis displays the LSR velocities based on the optical definition. An arrow in each panel indicates the recession velocity of the corresponding host galaxy. The galaxy name J0425$-$2521 in the bottom right panel refers to the galaxy 2MASXJ04250311$-$2521201 listed in Table \ref{tab:1}. The total isotropic H$_{2}$O maser luminosity for each detection is provided in Table \ref{tab:2}. }
\label{fig:detection}
\end{center} 
\end{figure*}

We detected new H$_{2}$O megamaser emission in four galaxies: CGCG 147$-$020, IRAS02438+2122, TGN229Z166, and 2MASXJ04250311$-$2521201 (hereafter J0425$-$2521). To enhance the SNR and better delineate the spectral profiles of the maser emissions, we applied boxcar smoothing to the spectrum of each detected source. The convolution kernel size used for smoothing was specified depending on the width of the maser line profile. Figure \ref{fig:detection} presents the boxcar-smoothed spectra for these four detections, with each panel showing an arrow that indicates the recession velocity of the galaxy. Table \ref{tab:2} summarizes the properties of these new detections, including the number of channels $n$ used for boxcar smoothing, the total isotropic maser luminosity $L_{\rm iso}$(H$_{2}$O), the peak flux density, the 1-$\sigma$ rms noise of the smoothed spectrum, and the SNR of the peak detection. The isotropic maser luminosity is calculated using an equation following \citet[e.g.][]{solomon05} :
\begin{equation}
L_{\rm iso}(\rm H_{2}O) = 1.04\times 10^{-3}~S(H_{2}O)\nu_{\rm rest}(1+z)^{-1}D_{\rm L}^{2} ~~L_{\odot},
\end{equation}
where S(H$_{2}$O) is the integrated maser line flux in units of Jy~km~s$^{-1}$, $\nu_{\rm rest}$ is the rest frequency of the maser line, $z$ is the redshift of the source, and $D_{\rm L}$ is the luminosity distance of the galaxy listed in Table \ref{tab:1}. For all cases, $L_{\rm iso}$(H$_{2}$O) $\gg$ 10 $L_{\odot}$, confirming that these are H$_{2}$O {\it megamasers} \citep[e.g.] []{kuo18}.

In these four megamaser spectra, we see no evidence of a {\it disk megamaser} profile (e.g. like the maser emission in NGC 4258) since their spectra do not show the characteristic ``triple-peaked profiles'', i.e. the spectra do not display three distinct maser line complexes that would correspond to the redshifted, systemic\footnote{The systemic masers refer to the maser spectral components having velocities close to the systemic velocity $v_{\rm sys}$ of the maser host galaxy.}, and blueshifted components of disk masers \citep[e.g.][]{kuo11, gao17}.  With the exception of J0425$-$2521, for which the SNR of the lines are too low for an in-depth exploration of the maser origin, the 22 GHz emission spectra only reveal a single broad maser line profile, with full width at half maximum (FWHM) of $\gtrsim 100-200$ km~s$^{-1}$. Given these characteristics, one could speculate that they may be like the {\it jet maser} systems similar to those hosted by NGC 1052 \citep{claussen98} and Mrk 348 \citep{peck02}, which exhibit a similar spectral profile. However, we note that the centroid ($v_{\rm c}$) of the maser line profile in a jet maser is often substantially redshifted or blueshifted from the systemic velocity ($v_{\rm sys}$) of the host galaxy \citep[e.g. $|v_{c}-v_{\rm sys}| \gtrsim 100$ km~s$^{-1}$;][]{peck02}, suggesting that these particular masers may originate from an off-nuclear cloud \citep[e.g. NGC 1068; ][]{gallimore96} that experiences a shock caused by the receding/approaching jet. Additionally, we also note that jet maser systems like NGC 1052 and Mrk 348 do not meet the mid-IR color criteria ($W1-W2 > 0.5$ and $W1-W4 > 7$), which differentiates them from our newly detected systems in terms of their mid-IR properties.

In contrast, all new maser detections from our GBT observations, each with a peak SNR $>$ 7, have maser line centroids or peaks lying very close to the systemic velocity ($v_{\rm sys}$) of their host galaxies (i.e., $|v_{c}-v_{\rm sys}| \lesssim 10$ km~s$^{-1}$). This suggests that a jet-cloud interaction in an off-nuclear region may not be the only explanation for the origin of this emission. The maser spectral profiles in these systems are similar to that of the unique gigamaser galaxy TXS 2226$-$184, which not only exhibits an exceptionally high maser luminosity of $L_{\rm iso}(\rm H_{2}O) = 6300$ $L_{\odot}$ \citep{kuo18}, but also has a maser line centroid close to $v_{\rm sys}$. Notably, this galaxy shares the same color criteria as our newly detected systems, with $W1-W2 > 0.5$ and $W1-W4 > 7$.

The latest VLBI observations of TXS 2226$-$184 reveal that H$_{2}$O masers in this system appear to be associated with the most luminous radio continuum clump of the nuclear region of the galaxy \citep{surcis20} and recent ALMA observations \citep{tarchi24} suggest that the maser emission in TXS 2226$-$184 likely arises from the amplification of a background continuum source by a foreground gas in a disk or torus \citep[e.g.,][]{gallimore96}. Given the similarities between TXS 2226$-$184 and our new maser detections, this scenario implies that these newly detected masers could share a similar origin as suggested by the ALMA observations, and we speculate that our mid-IR selection method may be particularly effective for identifying this class of maser systems. High-resolution VLBI mapping of these single-peaked maser systems would be essential to better characterize the morphology and origin of these maser activities.

With four detections out of the 77 observed galaxies, the overall maser detection rate in our GBT survey is 5.2\%. This rate is about 1.7 times higher than the maser detection rate of $\sim$3\% of the MCP survey, but significantly lower than the predicted detection rate of 18.2$\pm$2.5\% for red mid-IR type 2 galaxies that meet the color criteria
$W1-W2 > 0.5$ \& $W1-W4 > 7$.  A possible explanation is that the SED-fitting based AGN classification may not reliably identify true type 2 AGNs, leading instead to a sample of mid-IR red galaxies that is composed of a mix of galaxy types, including Type 1 AGNs and star-forming galaxies, rather than the type 2 AGNs targeted by the MCP survey. We further investigate this issue in the following section. 

\begin{deluxetable*}{lccccrrcc} 
\tablenum{2}
\tablewidth{0 pt} 
\tablecaption{Basic Properties of the Four New Megamaser Detections}
\tablehead{ 
\colhead{Galaxy} & \colhead{$t_{\rm int}$}  & \colhead{$T_{\rm sys}$} & \colhead{$n$}  &  \colhead{AGN}  & \colhead{$L_{\rm iso}$(H$_{2}$O)} & \colhead{ $F^{\rm peak}_{\nu}$ }  & \colhead{1 $\sigma$ rms} & \colhead{Peak} \\
\colhead{Name} & \colhead{(min)} & \colhead{(K)} & \colhead{(Channels)} &  \colhead{Type}   & \colhead{($L_{\odot}$)} & \colhead{(mJy)} & \colhead{(mJy)}  & \colhead{SNR} 
}     
\startdata 
CGCG 147$-$020         &  55  &  37.6  &  10  &  Sy2           &  764$\pm$3    & 20.4  & 1.7  &  12.0     \\   
IRAS 02438$+$2122      &  30  &  38.1  &  10  &  Liner         &  1881$\pm$96  & 35.4  & 2.3  &  15.4     \\ 
TGN 229Z166            &  10  &  51.8  &  50  &  Radio Galaxy  &  4467$\pm$10  & 22.4  & 2.7  &   8.3     \\
2MASXJ04250311$-$2521201 & 10 &  48.5  &  40  &  Sy2           &  1125$\pm$3   & 13.3  & 2.5  &   5.3     \\
\enddata
\tablecomments{Column (1): Galaxy name; Column (2): The on-source integration time of the observation; Column (3): The average system temperature of the observation; Column (4): Number of channels $n$ used for box smoothing; Column (5): AGN type of the maser galaxy. Except for TGN 229Z166, the AGN types are  from SIMBAD. For TGN 229Z166, the classification is adopted from \citet{sturm06}; Column (6): Total isotropic H$_{2}$O maser luminosity; Column (7): Peak flux density of the detection; Column (8): 1-$\sigma$ RMS noise of the maser spectrum after boxcar smoothing is applied; Column (9): Signal-to-noise ratio of the peak maser detection.}
\label{tab:2}
\end{deluxetable*}

\section{Discussion} \label{sec:discussion}

\subsection{The Impact of the Uncertainty in the SED-based Classification}

In general, the presence of water molecules suggests a dusty environment, indicating that the H$_{2}$O maser emission from an AGN may trace molecular material associated with a subparsec accretion disk or a molecular torus that obscures the central black hole, as proposed in the AGN unification scenario \citep{antonucci1985, kuo18}. Since the excitation of the 22 GHz H$_{2}$O maser emission also requires a warm environment \citep[i.e. a gas temperature $T_{\rm gas}\gtrsim 400$ K;][]{neu94, herrn05}, a heating source such as X-rays \citep[e.g.][]{nm95, kuo24} or shocks \citep[e.g.][]{hum08, dom15} is necessary to maintain $T_{\rm gas}$ within the preferred range for population inversion \citep[e.g.][]{lo05}, which could lead to enhanced mid-IR emission from the parsec-scale region of an AGN. Consequently, if a galaxy exhibits enhanced mid-IR emission, which often translates into a red infrared color, it may indicate the presence of an obscured AGN that could host an H$_{2}$O megamaser system.

\begin{figure*}[ht]
\begin{center} 
\vspace*{0 cm} 
\hspace*{0 cm} 
\includegraphics[angle=0, scale=0.6]{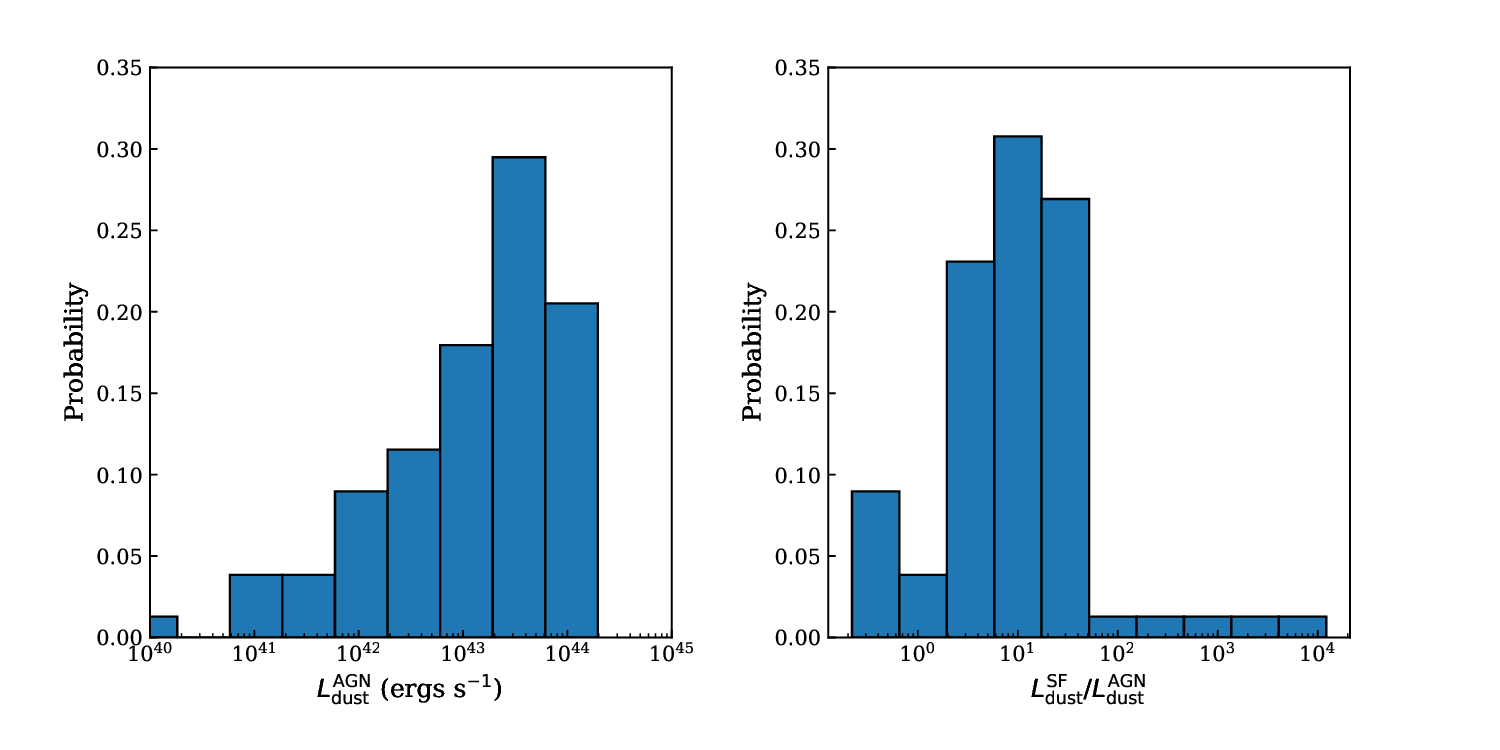}
\vspace*{0.0 cm} 
\caption{{\bf Left panel:} Probability distribution of the total IR luminosity of the AGN component in the parent sample of red galaxies, as obtained from SED-fitting. {\bf Right panel:} Probability distribution of the ratio of the total IR luminosity from star formation to that from the AGN. This plot illustrates that, for the majority of sources, infrared emissions from star formation dominates over those from the AGN component, i.e., $L^{\sc \rm SF}_{\rm dust}/L^{\sc \rm AGN}_{\rm dust} > 1$.}
\label{fig:SED}
\end{center} 
\end{figure*} 

For the sources of our GBT water maser survey, the presence of such a warm and dusty galactic nucleus is inferred from our SED-fitting, which suggests that the majority of our sources have a total infrared AGN luminosity $L^{\sc \rm AGN}_{\rm dust}\gtrsim 10^{42}$ ergs~s$^{-1}$ (see Figure \ref{fig:SED}), where $L^{\sc \rm AGN}_{\rm dust}$ refers to the total AGN luminosity in the $3-2000\ \micron$ infrared range. This IR luminosity range, along with the Seyfert 2 classification based on the SED best fits, strongly suggests an obscured AGN-like activity in the centers of these galaxies. Nevertheless, a close examination of these SED-fitting results suggests that the AGN identifications may carry significant uncertainty.  Our SED fits provide us with the total IR luminosity of both the AGN emission and that coming from star formation, and we compare these values for our sample in Figure \ref{fig:SED} (right panel).  We find that in most of our sources the IR emission is dominated by star formation activity rather than by the AGN, i.e., $L^{\sc \rm SF}_{\rm dust}/L^{\sc \rm AGN}_{\rm dust} > 1$ for the majority of the sample. As a result, our photometric data in the near-IR (i.e., $\lambda = 1.25, 1.65, 2.17 \micron$) and mid-IR wavelengths (i.e., $\lambda = 3.4, 4.6, 12, 22 \micron$) may not effectively distinguish between the SEDs of Type 1 and Type 2 AGNs, \citep[see Figure 2 in][]{chang17}, leading to potential misidentifications.

Therefore, our sample might not only include Seyfert 2 galaxies, but also Seyfert 1s and emission-line galaxies where activity is dominated by star formation. If this is the case, it is not surprising that the maser detection rate in our observation is significantly lower than what was predicted based on analysis of the mid-IR red galaxies of the MCP sample, where the H$_{2}$O megamasers are primarily detected in Seyfert 2 and LINER galaxies, and not in Seyfert 1s or other types of non-AGNs \citep[e.g.,][]{bra97}. 

To reliably investigate this possibility, we compiled optically-based AGN classifications for our sample from public extragalactic databases, including SIMBAD\footnote{\url{https://simbad.cds.unistra.fr/simbad/}} \citep{simbad}, which aggregates AGN types from various literature sources (see Column (9) in Table \ref{tab:1}). Notably, the majority of our newly detected maser emissions are found in Type 2 AGNs, consistent with previous water maser surveys. In contrast, our sample also contains a significant number of Seyfert 1 galaxies and emission-line galaxies dominated by star formation, which may account for the relatively low detection rate of masers.


\begin{deluxetable}{lcc} 
\tablenum{3}
\tablewidth{0 pt} 
\tablecaption{The AGN Classification Based on Optical-line Analysis }
\tablehead{ 
\colhead{Galaxy} & \colhead{Type}      &  \colhead{Type}  \\
\colhead{Name}   & \colhead{(Zaw19)}   &  \colhead{(simbad)}
}     
\startdata 
IC 883  &  T2-composite  &  Starburst  \\   
NGC 5610 & T1            &  Sy2 \\
TGN 121Z264 &  T1        &  Sy1 \\
UGC 04881NED01  &  T1    &  ---\\ 
NVSS J113639$+$173836    & T1  &  Radio Galaxy  \\
NVSS J130150$+$042001  & T1    &  Radio Galaxy \\
NGC 1377   &  Normal   &  --- \\
LEDA 966552  &  T2     &  Sy2\\
ESO 557-2  &  EmG      &  EmG\\
IRAS 11215$-$2806   & T2  & Sy2 \\
ESO 509$-$IG066NED01  &  T1  & Sy2\\
IRAS 13559$-$1553  & T2     & Sy2 \\
LEDA 90310  &   T1         &  Sy1\\
LEDA 855228  &  EmG        &  EmG \\
\enddata
\tablecomments{Column (1): Galaxy name; Column (2): AGN classification adopted from \citet{zaw19}. Here, T1 and T2 refer to Type 1 and Type 2 AGNs, respectively. "Normal" indicates a normal galaxy without emission lines. EmG denotes an emission-line galaxy dominated by star formation. T2-composite represents a system in which emission lines are significantly contributed by both star formation and the Type 2 AGN component; Column (3): AGN type from SIMBAD, listed here for comparison.}
\label{tab:3}
\end{deluxetable}

\subsection{The Estimation of Detection Rate for Type 2 AGNs}

To better evaluate the maser detection rate among optically selected Type 2 AGNs in our sample, we searched for accurate and secure AGN identifications for our sources. We find that our sample consists of a significant fraction (41\%) of galaxies from the 2MRS catalog \citep{huchra2012}, for which \citet{zaw19} provides the most complete and uniformly selected optical AGN catalog, using standard optical-line analysis. These classifications are the result of uniform spectral fitting to approximately 80\% of the 2MRS sources, providing thus a classical robust AGN identification. For a galactic plane cut of $|b| > 10^{\circ}$, that was imposed in order to improve the homogeneity of completeness for their sample, solid AGN identification was achieved for 61\% of the 2MRS catalog.  We list in Column (2) of Table \ref{tab:3} the AGN types for the fourteen 2MRS sources in our sample that are available from \citet{zaw19}.

As shown in Table \ref{tab:3}, our 2MRS subsample includes a substantial fraction (50\%) of Type 1 AGNs, while a smaller portion (21\%) consists of non-AGNs, which are a mix of normal and star-forming galaxies. Optically selected Type 2 AGNs, such as Seyfert 2s and LINERs, make up only 29\% of our 2MRS subsample. This finding supports our hypothesis that SED-based AGN classification may result in significant misidentifications when mid-IR emission from star formation overpowers that from the AGN in mid-IR luminous galaxies, which seems to be the case for our parent sample. The notable presence of non-Type-2 AGNs helps explain why the maser detection rate in our observations is substantially lower than the expected rate of 18.2$\pm$2.5\% from the MCP survey \citep{kuo18}, which primarily targets Seyfert 2s and LINERs. 
However, if we consider only the securely classified Type 2 AGNs, the maser detection rate would be either 17.9\% or 13.4\%, depending on whether the maser detection TGN229Z166 is classified as a Type 2 AGN. Although TGN229Z166 exhibits emission-line AGN characteristics, as noted by \citet{sadler02}, its specific subtype remains poorly defined in the literature, adding uncertainty to our detection rate estimate. Nonetheless, the revised maser detection rate among Type 2 AGNs is consistent (within 1$-$2 $\sigma$) with the expected detection rate calculated for MCP galaxies that meet the color criteria $W1-W2 > 0.5$ and $W1-W4 > 7$. This further supports the conclusion that the maser detection rate is significantly higher in mid-IR red galaxies hosting Type 2 AGNs, as proposed by \citet{kuo18}.



\section{Conclusion} \label{sec:conclusion}

In this study, we utilized new GBT observations to evaluate whether selecting mid-IR red galaxies leads to a high maser detection rate, as predicted \citep[i.e., $\sim$18\%;][]{kuo18}.   
Our observations resulted in the detection of H$_{2}$O maser emission in 4 out of 77 mid-IR red galaxies with WISE colors satisfying $W1-W2 > 0.5$ and $W1-W4 > 7$, for which SED fitting suggested a Seyfert 2 origin of their dominant optical emission, yielding a detection rate of 5.2\%. We note that this lower-than-expected detection rate most likely results from the fact that SED-fitting may carry significant uncertainty in distinguishing between the different types of AGNs, especially when the IR emission from starformation dominates over that from AGN. As a result, the predicted high water maser detection rate does not apply to the initial parent sample of mid-IR red galaxies, as it does not match the parent sample that was used to derive the $\sim$18\% detection rate by \citet{kuo18} in the first place. Our analysis is likely the first to quantitatively demonstrate that, while MAGPHYS can effectively identify IR-luminous AGNs, it may have limitations in differentiating between Seyfert 1 and Seyfert 2 galaxies.

Nevertheless, we show that, by estimating the level of potential contamination by non-Type-2 AGNs, the water maser detection rate aligns well with the expected value if only the subset of optically classified Type 2 AGNs were considered. This finding reinforces the idea that the maser detection rate is significantly higher in mid-IR red galaxies that host Type 2 AGNs \citep{kuo18}. In addition, it also suggests that the mid-IR selection is most effective if one starts with optically-selected Type-2 AGNs in a maser survey, followed by applying the mid-IR cutoffs. 

With the identification of $\gtrsim 6000$ new Type 2 AGNs from the 2MRS catalog within $z \lesssim 0.05$ \citep{zaw19} which have not yet been surveyed for water megamasers, and the potential for discovering many more Type 2 AGNs in the ongoing DESI dark energy survey \citep[e.g.,][]{DESI24}, it is promising that one can still detect $\gtrsim$180 H$_{2}$O megamaser systems (i.e. $\sim$3\% of $\gtrsim$6000) in the future. Given that $\sim$6\% (or $\sim$360) of the new $\gtrsim$6000 Type 2 AGNs would meet the color criteria $W1-W2 > 0.5$ and $W1-W4 > 7$ (see Section \ref{sec:sample_selection}), it is expected that $\sim$65 of the $\gtrsim$180 megamasers (i.e. 18\% of 360) existing in the new Type 2 AGNs can be effectively identified with the mid-IR selection method presented in this paper. Nevertheless, the absence of disk maser systems in our detections, potentially due to the small sample size of Type 2 AGNs, leaves it unclear whether mid-IR selection can also effectively identify disk maser systems, which are crucial for accurate black hole mass and Hubble constant measurements. Further testing with a larger survey of optically identified Type 2 AGNs is required to confirm the effectiveness of mid-IR selection for detecting disk masers.

\section*{Acknowledgements}

This publication is supported by National Science and Technology Council, R.O.C under the project 113-2112-M-110 -013-. A.C. acknowledges support from the National Science Foundation under Grant No. AST 1814594. The National Radio Astronomy Observatory and Green Bank Observatory are facilities of the U.S. National Science Foundation operated under cooperative agreement by Associated Universities, Inc. This research has made use of NASA's Astrophysics Data System Bibliographic Services, and the NASA/IPAC Extragalactic Database (NED) which is operated by the Jet Propulsion Laboratory, California Institute of
Technology, under contract with the National Aeronautics and Space Administration. In addition, this work also makes use of the cosmological calculator described in \citet{wright06} as well as the SIMBAD database,
operated at CDS, Strasbourg, France \citep{simbad}.

\bibliography{references}{}
\bibliographystyle{aasjournal}



\end{document}